\documentclass[reprint,twocolumn,prl]{revtex4}
\usepackage{amssymb}
\usepackage[T1]{fontenc}
\usepackage[latin9]{inputenc}
\usepackage{color}
\usepackage{amsmath}
\usepackage{graphicx}
\usepackage{esint}
\usepackage{bm}

\newcommand{\diag}{\operatorname{diag}}
\newcommand{\Real}{\operatorname{Re}}

\makeatletter
\@ifundefined{textcolor}{}
{ \definecolor{BLACK}{gray}{0}
 \definecolor{WHITE}{gray}{1}
 \definecolor{RED}{rgb}{1,0,0}
 \definecolor{GREEN}{rgb}{0,1,0}
 \definecolor{BLUE}{rgb}{0,0,1}
 \definecolor{CYAN}{cmyk}{1,0,0,0}
 \definecolor{MAGENTA}{cmyk}{0,1,0,0}
 \definecolor{YELLOW}{cmyk}{0,0,1,0}
 }

\begin{document}

\title{Local-realistic Bohmian trajectories: a non-Bohmian approach to wave-particle duality}
\author{F. De Zela}

\affiliation{Departamento de Ciencias, Secci\'{o}n F\'{i}sica, Pontificia Universidad Cat\'{o}lica del Per\'{u}, Lima 15088, Peru}




\begin{abstract}

We present a local-realistic description of both wave-particle duality and Bohmian trajectories. Our approach is relativistic and based on Hamilton's principle of classical mechanics, but departs from its standard setting in two respects. First, we address an ensemble of extremal curves, the so-called Mayer field, instead of focusing on a single extremal curve. Second, we assume that there is a scale, below which we can only probabilistically assess which extremal curve in the ensemble is actually realized. The continuity equation ruling the conservation of probability represents a subsidiary condition for Hamilton's principle. As a consequence, the ensemble of extremals acquires a dynamics that is ruled by Maxwell equations. These equations are thus shown to also rule some non-electromagnetic phenomena. While particles follow well-defined trajectories, the field of extremals can display wave behavior.

\end{abstract}

\maketitle


Bohmian trajectories (BT) represent a particle's feature that can be derived from Schr\"{o}dinger's wave equation. Once dubbed as ``surreal'', such a labeling has been rendered inappropriate by recent experiments  \cite{kocsis2011,mahler2016,Xiao2017}, in which single-photon BT were registered through weak-value measurements \cite{aharonov}. There were two main reasons to see BT as surreal. First,
the orthodox (Copenhagen) interpretation of quantum mechanics (QM) forbids assigning physical reality to particle trajectories, as these presuppose well-defined positions and velocities. Second, BT were numerically calculated \cite{philippidis} and experimentally measured for a two-slit Young setup, in which they displayed curvilinear motion. BT are therefore at odds with the principle of inertia \cite{cohen2020}.
To be sure, the physical interpretation of BT is still open to debate \cite{englert1992,scully1998,dewdney1993,durr1993,englert1993,wiseman2003,schleich2013,wu2013,bliokh,jooya2016,Tastevin2018,yu2018,cheng2018,douguet2018,Xiao2019,li2021}, but the very debate shows that the Copenhagen interpretation has not really established itself as a full-fledged paradigm that replaced the classical one. The quantum formalism can actually be used without adhering to the Copenhagen interpretation. Perhaps most physicists keep thinking of an electron as a well-localized particle that moves with a well-defined velocity, thereby adopting a realist view as long as it is not necessary to deal with foundational issues. The latter have lately sparked interest in a community that includes application-oriented scientists. This is because the so-called ``quantum advantage'' is claimed to derive from unique quantum features, such as coherent state-superposition and entanglement. It is thus important to make sure that it is impossible to employ conventional technology, which is based on classical physics, in order to implement quantum algorithms in an efficient and scalable way. Impossibility claims should however rest on a firm scientific ground rather than on ideological views.

Wave-particle duality (WPD) is also seen as a distinctive quantum feature. Indeed, Feynman famously said \cite{feynman} that WPD is a phenomenon that ``is impossible, \emph{absolutely} impossible, to explain in any classical way''. Feynman's claim does not rest on a firm, scientific basis. Not even Bell's theorem \cite{mermin1993} could provide such a basis for a claim that is representative of a philosophical view rather than a demonstrable proposition. Said view arose in a cultural environment that was hostile to rationalism and causality in general, including their manifestation in classical physics \cite{meyenn}. The extraordinary scientific stature of people like Bohr, Heisenberg, Pauli, Dirac, Born and other developers of QM, contributed much to disseminate their philosophical view and to make it the prevailing one. It is by now common place to say that WPD ascribes mutually exclusive properties to one and the same physical object. We are, however, not compelled to subscribe such a view \cite{selleri,aharonov2017}. We may ascribe particle-like properties to an electron, say, while wave-like properties characterize the \emph{probability} to find the electron at a given place and time. Electron and probability are two different concepts. One of them may be characterized by particle features, and the other by wave features. Only particular, historical circumstances that defined the European environment of the early twentieth century led physicists to see a paradox where there is none. In economics, for instance, nobody sees a paradox in establishing mathematical -- deterministic or stochastic -- models for some commodities market, while recognizing the unpredictable, free-will behavior of the individuals who drive this market.

There are two main lines of research regarding WPD. One of them is concerned with its quantification and started in 1979 with the work of Wootters and Zurek \cite{wooters}, which gave rise to much development in this field  \cite{wooters,scully1991,jaeger,englert1996,jakob2010,Coles2014,qian2016,eberly2017,qian2018,norrman2020,miranda2021,qureshi2021,qian0}. The other research line started in the mid-1920s and was concerned with a realistic interpretation of the Schr\"{o}dinger equation. A first attempt was that of Madelung \cite{holland1993}, which was further developed by Bohm \cite{bohm} and others. This line of research led to the BT that we address here. We will present an alternative approach to that of Madelung, de Broglie and Bohm  \cite{holland1993}, an approach that fully fits within the classical framework. In particular, we present a local-realistic description of the two-slit experiment. We do this not for the sake of proving Feynman wrong, but to explore possible new avenues towards a better understanding of the quantum-classical boundary \cite{haroche2013,nassar2013}.
Schr\"{o}dinger's equation shows how probability, a rather abstract and -- in its Bayesian formulation -- subjective concept, can nonetheless be subjected to deterministic dynamics, as though it were a physical object. We show here that classical physics admits a similar treatment. Indeed, from Hamilton's principle and the conservation of probability, one can derive Maxwell-like equations that rule the motion of a particle and its associated probability density.

\section*{Bohmian mechanics and classical optics}
As is well-known, de Broglie prompted Schr\"{o}dinger to find his equation. Schr\"{o}dinger's equation was intended to be for material particles what the wave equation was for optical phenomena. The optical wave equation can be derived from Maxwell equations, the short-wave limit of which leads to ray optics. The latter can alternatively be based on Fermat's principle, which is the optical counterpart of Hamilton's principle in mechanics. In his quest to find the equivalent of the optical wave equation in mechanics, Schr\"{o}dinger postulated his equation. Its precise link to classical mechanics remains undetermined.

Let us first discuss the connection between the optical wave equation and Schr\"{o}dinger's equation.
In scalar wave optics \cite{saleh,wolf}, light propagation in vacuum is described by a real function
$u(\mathbf{r},t)$, which satisfies $(\partial^2/c^2\partial t^2-\nabla^2)u=0$. The optical intensity is given by $I(\mathbf{r},t)=2\langle u^2(\mathbf{r},t)\rangle$, where angular brackets denote averaging over a time much longer than any optical cycle. In the monochromatic case, it is convenient to introduce a complex function $U(\mathbf{r},t)=U(\mathbf{r})e^{i\omega t}$, such that $u(\mathbf{r},t)=\Real U(\mathbf{r},t)$. Then, $I(\mathbf{r})=|U(\mathbf{r})|^2$. If light propagates in a medium of refractive index $n(\mathbf{r})$, the wave equation reads
\begin{equation}\label{weq2}
 \left(\frac{1}{v^2}\frac{\partial^2}{\partial t^2} -\nabla^2\right)U(\mathbf{r},t)=0,
\end{equation}
with $v=c/n$. On setting $U(\mathbf{r},t)=U(\mathbf{r})e^{i\omega t}$ in Eq. (\ref{weq2}), we get the Helmholtz equation
\begin{equation}\label{helmholtz}
  \nabla^2 U(\mathbf{r})+k^2 U(\mathbf{r})=0,
\end{equation}
with $k=n \omega/c$.  Let us set $U(\mathbf{r})=a(\mathbf{r})\exp[-ik_0 \tilde{S}(\mathbf{r})]$, with $k_0=\omega/c$, the wave number in vacuum. Separating real and imaginary parts of Eq. (\ref{helmholtz}) yields
\begin{eqnarray}
  |\nabla \tilde{S}|^2-n^2-\lambdabar^2 \frac{\nabla^2 a}{a}  &=& 0 \label{realopt}\\
  2\boldsymbol{\nabla}a \cdot \boldsymbol{\nabla}\tilde{S}+a\nabla^2 \tilde{S} &=&0, \label{imagopt}
\end{eqnarray}
where $\lambdabar=1/k_0$. If $a(\mathbf{r})$ varies slowly over distances in the order of $\lambdabar$, we can neglect the second term on the right-hand side of Eq.~(\ref{realopt}), thereby obtaining the eikonal equation: $|\nabla \tilde{S}|^2 = n^2$. This is the domain of ray optics. Rays are defined as trajectories orthogonal to the level surfaces $\tilde{S}(\mathbf{r})=\text{const}$. Setting the arc-length $s$ as curve parameter, light rays are solutions of the differential equation
\begin{equation}\label{rays}
  \frac{d\mathbf{r}}{ds}=\frac{1}{n(\mathbf{r}(s))}\boldsymbol{\nabla}\tilde{S}(\mathbf{r}(s)).
\end{equation}

Let us now turn to the Schr\"{o}dinger equation: $i\hbar \partial \psi/\partial t=-\hbar^2\nabla^2 \psi/2m+V(\mathbf{r})\psi$. Setting $\Psi(\mathbf{r},t)=R(\mathbf{r},t)\exp{(i S(\mathbf{r},t)/\hbar)}$ and splitting real and imaginary parts gives
\begin{eqnarray}
\frac{\partial S}{\partial t}+\frac{|\nabla S|^2}{2m}+V -\frac{\hbar^2}{2m}\frac{\nabla^2 R}{R}&=& 0, \label{realsc2}\\
  \frac{\partial R^2}{\partial t}+\boldsymbol{\nabla}\cdot\left(R^2\frac{\boldsymbol{\nabla}S}{m}\right)&=&0. \label{imagsc2}
\end{eqnarray}
For $R=R(\mathbf{r})$ and $S(\mathbf{r},t)=-Et +S(\mathbf{r})$, the above equations reduce to Eqs.~(\ref{realopt}) and (\ref{imagopt}), with the replacements $S/\hbar \rightarrow \tilde{S}/\lambdabar$, $\hbar R^2/m \rightarrow \lambdabar a^2$ and $2m(E-V)/\hbar^2 \rightarrow n/\lambdabar^2$. Eq.~(\ref{realsc2}) is the Hamilton-Jacobi equation with the ``quantum potential'' $Q=-\hbar^2\nabla^2 R/(2mR)$ added to $V$. Eq.~(\ref{imagsc2}) is a continuity equation for the probability density $\rho=R^2$ and probability current $\mathbf{j}=R^2 \boldsymbol{\nabla}S/m$. From $S(\mathbf{r})$, one obtains particle trajectories by integrating
\begin{equation}\label{rays2}
  \frac{d\mathbf{r}}{dt}=\frac{1}{m}\boldsymbol{\nabla}S(\mathbf{r}(t)).
\end{equation}
Eqs.~(\ref{realsc2}), (\ref{imagsc2}) and (\ref{rays2}) are the basis of Bohm's reformulation of Schr\"{o}dinger's wave mechanics.
Finding $S(\mathbf{r})$ in the case of Young's double-slit configuration is a difficult task. A more viable approach is to get $S$ and $R$ from a solution $\Psi(\mathbf{r})=R(\mathbf{r})\exp{(i S(\mathbf{r})/\hbar)}$ of the time-independent Schr\"{o}dinger equation for free particles and double-slit boundary conditions. On using Feynman's path integral method, Philippidis \textit{et al}. \cite{philippidis} numerically obtained BT with data taken from experiments performed by J\"onsson \cite{jonsson}. It is important to notice that these experiments did not test the Schr\"{o}dinger equation itself, but a consequence of it, Helmholtz's equation (\ref{helmholtz}).

Bohm's approach provides a realistic and deterministic framework whose predictions coincide with those of QM. However, the price paid for this deterministic version is a drastic departure from the most basic tenets of classical physics. Particles following BT display curvilinear motion, even though there is no external force acting on them. Curvilinear motion is caused by $Q$, the quantum potential. This potential, besides being non-local, stems from the particle itself through its associated ``pilot-wave''. This amounts to endow probability (amplitudes) with physical existence. Hence, by adopting Bohm's approach, we must accept self-action as a matter of principle, something that is even more at odds with the basic tenets of classical physics than QM itself.

No departure from the classical formalism is needed to explain J\"onsson's interference patterns and BT. A classical description may include both particle-like features and wave-like features. In classical mechanics, particles move along extremal curves of an action principle: $\delta \int L(x,\dot{x})ds=0$. An extremal curve is a solution of the Euler-Lagrange equations for the Lagrangian $L(x,\dot{x})$. The variational problem, which consists in finding an extremal curve, in fact requires finding a whole set of extremals. The sought-after extremal curve must be in fact a member of a whole field of such curves. This is the imbedding theorem in the calculus of variations \cite{bliss,Caratheodory,Rund}, basically a consequence of continuity assumptions.

\begin{figure}[ht!]
\centering
\fbox{\includegraphics[width=\linewidth]{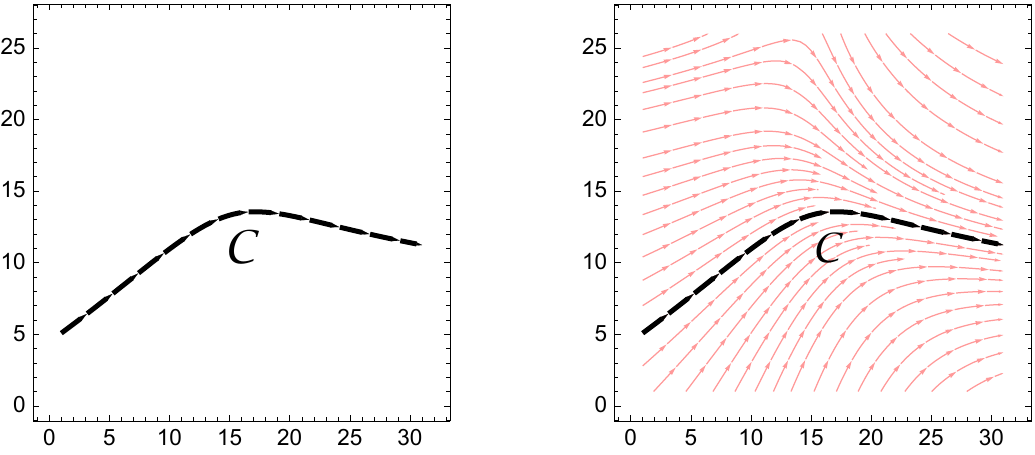}}
\caption{Left panel: curve $C$ renders $\int L(x,\dot{x})ds$ extremal. One usually focuses on a single extremal curve, even though this curve must be embedded in a whole family of extremals. Right panel: the complete picture. $C$ is just one member of a whole family of curves, whose collective behavior is ruled by field equations.}
\label{fig3}
\end{figure}

The left panel of Fig.~(\ref{fig3}) illustrates the usual approach, in which a single extremal is addressed. The right panel shows instead the whole picture: the sought-after extremal exists only within a Mayer field \cite{bliss}, a ``pilot-field'', as we may call it. Depending on the boundary conditions, one and the same Lagrangian can lead to different fields. A screen with two open slits is not the same as a screen with only one open slit. A particle that goes through one slit can ``know'' whether the other slit is open or not, because its trajectory belongs to a field whose structure depends on whether the two slits are open or not. While it hardly makes sense to say that a single particle interferes with itself, a field may well develop wave dynamics. Interference phenomena are thus possible for fields of extremals. A two-slit scenario can lead to an interference pattern. By sending and detecting one particle after the other, we can exhibit such a pattern. There is no conflict between the two features that show up here: particle and wave may coexist.

\section*{Carath\'{e}odory's ``royal road''}

The foregoing statements derive from Hamilton's principle: $\delta I \equiv \delta \int L(x,\dot{x})ds=0$.
The existence of the Mayer field imposes some integrability conditions. Carath\'{e}odory's ``royal road'' to the Calculus of Variations \cite{Caratheodory}
leads at once to Euler-Lagrange, Hamilton and Hamilton-Jacobi equations.

Let us summarize Carathéodory's approach, focussing on a relativistic formulation, whose action principle has the form $\delta \int L(x,\dot{x})ds=0$. In Carathéodory's approach, $L(x,\dot{x})$ is written as a function of $x$ and the velocity-field $v(x)$: $L=L(x,v(x))$. Extremal curves $\mathcal{C}$ are solutions of
\begin{equation}\label{ic}
  \frac{dx^{\mu }(s)}{ds}=v^{\mu}(x(s)).
\end{equation}
We seek for a field $v(x)$ whose integral curves render $\int Lds$ extremal. To this end, we introduce an auxiliary function $S(x)$ and require that the following, fundamental equations are satisfied \cite{Caratheodory,Rund,supp}:
\begin{eqnarray}
 L(x,v)-v^{\mu }(x)\partial_{\mu }S(x)&=&0 \label{c1}\\
  \partial L(x,v)/\partial v^{\mu}&=&\partial S/\partial x^{\mu}.\label{c2}
\end{eqnarray}
On account of the integrability conditions $\partial ^{2}S/\partial x^{\mu }\partial x^{\nu }=\partial
^{2}S/\partial x^{\nu }\partial x^{\mu }$, it follows from (\ref{c2}) that
\begin{equation}
\frac{\partial }{\partial x^{\mu }}\left( \frac{\partial L(x,v(x))}{\partial v^{\nu }}%
\right) -\frac{\partial }{\partial x^{\nu }}\left( \frac{\partial L(x,v(x))}{\partial
v^{\mu }}\right) =0.  \label{integrability}
\end{equation}%
From Eq.~(\ref{integrability}), we can get the equations of motion \cite{supp}. Carath\'{e}odory's formulation addresses local properties, rather than some particular extremal curve. The latter can be singled out by choosing initial values $x^{\mu}(s_0)=x^{\mu}_0$ when solving Eq.~(\ref{ic}). We will consider situations for which $x^{\mu}_0$ occurs with some probability and introduce a normalized probability distribution $\rho(x^{\mu})\geq 0$, with $\int \rho(x^{\mu})d^4 x=1$. Notice that this is not the standard framework of statistical mechanics, where $\rho$ depends on both position and momentum: $\rho(x^{\mu},p_{\mu})$.

\section*{Two local-realistic descriptions}

Let us consider now a free particle. Its Lorentz invariant Lagrangian reads
\begin{equation}\label{lagrangianb}
    L(x,v)=mc(\eta _{\mu \nu }v^{\mu }v^{\nu })^{1/2}\equiv mc \, \phi,
\end{equation}
where $\eta _{\mu \nu }=\diag(+1,-1,-1,-1)$ is the Minkowski metric tensor. It holds $\partial L/\partial v^{\nu}=m c v_{\nu}/\phi=\partial S/\partial x^{\nu}$. Given $v^{\nu}$, we can find a $\rho(x)$, such that $\partial_{\nu}(\rho v^{\nu})=0$. Indeed, this ``continuity equation'' can be written in the form $v^{\nu}\partial_{\nu}\ln\rho = -\partial_{\nu}v^{\nu}$, which has a solution $\rho(x)$ for fairly general boundary conditions. Moreover, the integrability conditions $\partial^2 S/\partial x^{\mu}\partial x^{\nu}=\partial^2 S/\partial x^{\nu}\partial x^{\mu}$ lead to
\begin{equation}\label{n1}
  \partial_{\mu}w_{\nu}-\partial_{\nu}w_{\mu}=0,
\end{equation}
with $w_{\nu}=v_{\nu}/\phi$. From $w^{\mu}w_{\mu}=1$, it follows that $w^{\mu}\partial_{\nu}w_{\mu}=0$. Thus,
Eq.~(\ref{n1}) implies $w^{\mu}\partial_{\mu}w_{\nu}=0$, which means that $d w_{\nu}/ds=0$, i.e, the extremals are straight lines. This corresponds to an idealized description, in which measurement devices have infinite resolution and particles have exact, sharply defined locations $x^{\mu}$.

\subsection*{First probabilistic model}
Assume now that there is a scale, below which we can assign locations only probabilistically. Both $v^{\nu}(x)$ and $\rho(x)$ must now be determined together. Assume further that $\rho= n_0c/\phi$, with $n_0$ a constant providing dimensional consistency. $\rho(x)$ is thus given by the vector-norm $v_{\nu}(x)v^{\nu}(x)$, a possible quantifier for the ``density'' of the field-lines.
Eq.~(\ref{n1}) now reads $\partial_{\mu}(\rho v_{\nu})-\partial_{\nu}(\rho v_{\mu})=0$. Using $\partial^{\mu}(\rho v_{\mu})=0$, we get
$\partial^{\mu}\partial_{\mu}(\rho v_{\nu})-\partial^{\mu}\partial_{\nu}(\rho v_{\mu})=\partial^{\mu}\partial_{\mu}(\rho v_{\nu})\equiv \square (\rho v_{\nu})=0$. Hence, we seek for $v^{\nu}$, such that
\begin{equation}\label{n2}
  \square (\rho\, v_{\nu})=0, \quad \text{and} \quad \partial^{\mu}(\rho \, v_{\mu})=0,
\end{equation}
with $\rho=n_0c(v_{\nu}v^{\nu})^{-1/2}$. Finding such a $v^{\nu}$ is generally a difficult task. We can deal instead with the probability current
\begin{equation}\label{pc}
  \pi^{\nu}(x)=\rho(x) v^{\nu}(x),
\end{equation}
the equations for which read
\begin{equation}\label{n3}
  \square \pi^{\nu}=0, \quad \text{and} \quad \partial_{\nu}\pi^{\nu}=0.
\end{equation}
These are identical to the source-free Maxwell equations in the Lorentz gauge. Indeed,
let us define the antisymmetric tensor
\begin{equation}\label{n4}
M_{\mu \nu}= \partial_{\mu}\pi_{\nu}-\partial_{\nu}\pi_{\mu}.
\end{equation}
On view of Eqs.~(\ref{n3}) and (\ref{n4}), $M_{\mu \nu}$ satisfies the Maxwell-like equations
\begin{eqnarray}
    \partial _{\mu }M^{\mu \nu}&=&0, \label{n5a} \\
     \partial_{\alpha}M_{\beta\gamma}+\partial_{\beta}M_{\gamma \alpha}+\partial_{\gamma}M_{\alpha\beta}&=&0, \label{n5b}
\end{eqnarray}
where (\ref{n5b}) is an identity implied by (\ref{n4}).
Eqs.~(\ref{n5a}) and (\ref{n5b}) show that the velocity field of a nominally ``free'' particle can be ruled by Maxwell-like equations. This comes from supplementing the equations of motion with the continuity equation. One may wonder how conservation of probability can have dynamical consequences, so that nominally ``free'' particles do not necessarily move along straight lines. We return to this question later. Eqs.~(\ref{n5a}) and (\ref{n5b}) have a great number of known solutions, i.e., those which have been obtained in electrodynamics. Only some of them will have physical meaning in our case. Given $\pi^{\nu}$, we can obtain $v^{\nu}$ from $\rho v^{\nu}=\pi^{\nu}$, viz., $n_0c\, v^{\nu}=(v_{\sigma}v^{\sigma})^{1/2}\pi^{\nu}$, with $\nu=0,\dots, 3$. By squaring these four equations, we readily see that, for a time-like $\pi^{\nu}$, with $\pi_{\sigma}\pi^{\sigma}=n_0^2 c^2$, there are infinitely many $v^{\nu}$, with one of the $v^{\nu}$ being a free parameter \cite{supp}.

Setting $\pi^{\nu}(x)\equiv n(x) v^{\nu}(x)$, the first of Eqs.~(\ref{n3}) also follows from the Lagrangian
$L^{\prime}(x,v)=(n(x)v_{\nu}v^{\nu})^{1/2}$. Indeed, the condition $\partial_{\nu}(n v^{\nu})=0$ leads here again to the equations $\square(n v^{\nu})=0$, provided we choose the normalization $(v_{\nu}v^{\nu})^{1/2}=1$, which is always possible \cite{supp}. $L^{\prime}$ and its associated Hamilton's principle, $\delta \int L^{\prime}(x,\dot{x})ds=0$, are relativistic generalizations of Fermat's principle in optics, which involves the refractive index of a background medium. This suggests interpreting $\rho=n_0c(v_{\nu}v^{\nu})^{-1/2}$ as a probability density that also carries information of the background medium, in this case the electromagnetic (EM) vacuum. Such a medium can have physical properties that, under appropriate circumstances, may affect the motion of ``free'' particles. The presence of the EM vacuum in our description can be exposed by writing $c=(\epsilon_{0}\mu_{0})^{-1/2}$, where $\epsilon_{0}$ is the electric permittivity and $\mu_{0}$ the magnetic permeability. While these considerations are rather speculative, they fully fit into the classical framework. We may recall that $c$, with its purely EM content, also appears in various equations that are purported to describe non-EM phenomena.
Likewise, Planck's constant, the single known candidate for setting a scale for the quantum-classical boundary, is a purely EM quantity: $\hbar=137.036 (\epsilon_0 \mu_0)^{1/2}e^2$, where $e$ is the electron's charge. In retrospect, it was a most unfortunate decision to include $c$ and $\hbar$ among the ``fundamental constants''. As a consequence of this decision, some natural questions remained unasked. For instance, consider Schwarzschild's solution ($g_{\mu\nu}$) of Einstein's equations. It reads $g_{\mu\nu}dx^{\mu}dx^{\nu}=(A(r)/\epsilon_0\mu_0)dt^2-(1/A(r))dr^2-r^2d\Omega^2$, where $A(r)=1-2 G M \epsilon_0\mu_0/r$. This immediately begs the question: why do electromagnetic properties enter a purely gravitational effect? Even the event-horizon radius $r_S$ of a black hole, given by $A(r_S)=0$, depends on $\epsilon_0\mu_0$.
Consider next neutrino oscillations between two neutrino types, electron- and muon-neutrino. The oscillation probability is given by
$P_{\nu_e \rightarrow\nu_{\mu}}=\left[\sin(2\theta)\sin(\Delta E \, t/2\hbar)\right]^2$. On setting
$\hbar=137.036 (\epsilon_0 \mu_0)^{1/2}e^2$ in this formula, we are led to ask: why does the electron's charge show up in a process that involves only neutral particles?

\subsection*{Second probabilistic model}

If we see the EM vacuum as a medium whose physical properties are characterized by $\epsilon_{0}$ and $\mu_{0}$, we can go a step further and assume that this medium provides a causal
connection between $\pi^{\mu}(x^{\prime})$ and $\pi^{\mu}(x)$, the probability current densities at two space-time points. In consonance with Eq.~(\ref{n3}), we assume that said connection propagates as prescribed by a Green function $G(x)$ that satisfies $\square_x G(x-x^{\prime})=\delta^{(4)}(x-x^{\prime})$:
\begin{equation}\label{jj}
  \pi^{\mu}(x)=\kappa \int G(x-x^{\prime})\pi^{\mu}(x^{\prime})d^{4}x^{\prime}.
\end{equation}
$\kappa$ is a constant that makes Eq.~(\ref{jj}) dimensionally correct: it has units of inverse-length squared. This length sets a scale in our description, similarly to $\hbar$ in QM. Notice that Eq.~(\ref{jj}) is in line with similar descriptions in classical physics, such as linear-response theory, scattering theory, etc. As an example, consider a plane, monochromatic wave that propagates along the $z$-axis. Writing $\boldsymbol{\rho}=(x,y)$, the transverse electric field $E_{j=x,y}(\boldsymbol{\rho},z,\omega)$ on plane $z$ is given by \cite{wolf}
\begin{equation}\label{wolf}
 E_{j}(\boldsymbol{\rho},z,\omega)=e^{ikz}\int_{(z=0)}G(\boldsymbol{\rho}-\boldsymbol{\rho}^{\prime},z;\omega)E_{j}(\boldsymbol{\rho}^{\prime},0,\omega)d^{2}\boldsymbol{\rho}^{\prime},
\end{equation}
where $G(\boldsymbol{\rho}-\boldsymbol{\rho}^{\prime},z;\omega)$ is a Green function for paraxial propagation. Eq.~(\ref{jj}) can also be seen as a generalized Huygens-Fresnel principle: $\pi^{\mu}(x^{\prime})$ produces a disturbance that propagates as prescribed by $G(x)$ and, as a result, we have $\pi^{\mu}(x)$. More precisely, this means the following. If a particle happens to be at $x^{\prime}$ and moving with velocity $v^{\mu}(x^{\prime})$, it would cause that, some time later, in the eventuality that a particle is at $x$, it will move with velocity $v^{\mu}(x)$. Formulated in terms of the corresponding probabilities, we have Eq.~(\ref{jj}). Causality is assured by choosing a ``retarded'' Green function. This function, in turn, transcribes the properties of the background medium, in our case EM vacuum.

From Eq.~(\ref{jj}), we get
\begin{equation}\label{kg2}
  \square \pi^{\mu}=\kappa \, \pi^{\mu}.
\end{equation}
Under Dirichlet or von Neumann boundary conditions, we can readily obtain \cite{supp}
\begin{equation}\label{k1}
  \kappa=-\frac{\int_{V}(\partial_{\sigma}\pi_{\mu})(\partial^{\sigma}\pi^{\mu})dV}{\int_{V}\pi_{\mu}\pi^{\mu}dV}.
\end{equation}
While we may assume that $\pi^{\mu}$ is time-like ($\pi_{\mu}\pi^{\mu}>0$), the numerator of Eq.~(\ref{k1}) can be positive or negative. Let us take $\kappa <0$. Setting $\kappa=-1/\lambda^{2}$, we can write Eq.~(\ref{kg2}) as a Proca-type equation:
\begin{equation}\label{kg6}
  \left(\square +\frac{1}{\lambda^2}\right) \pi^{\mu}(x)=0.
\end{equation}
From Eq.~(\ref{kg6}), we can get BT. Before showing this, we derive here again
Maxwell-type equations for
\begin{equation}\label{tensork}
K^{\alpha \beta}=\partial^{\alpha}\pi^{\beta}-\partial^{\beta}\pi^{\alpha}.
\end{equation}
From $\partial_{\alpha}K^{\alpha\beta}=\partial_{\alpha}\partial^{\alpha}\pi^{\beta}-\partial^{\beta}(\partial_{\alpha}\pi^{\alpha})$, on account of $\partial_{\alpha}\pi^{\alpha}=0$ and Eq.~(\ref{kg2}), we get
\begin{equation}\label{max}
  \partial_{\alpha}K^{\alpha\beta}=\kappa \, \pi^{\beta},
\end{equation}
which are formally identical to the non-homogeneous Maxwell equations. The homogeneous equations follow from
Eq.~(\ref{tensork}), as an identity:
\begin{equation}\label{max2}
    \partial^{\alpha}K^{\beta\gamma}+\partial^{\beta}K^{\gamma \alpha}+\partial^{\gamma}K^{\alpha\beta}=0.
\end{equation}
Hence, reduced to the bare essentials, the above Maxwell-like equations reflect nothing but propagation properties, namely those encoded in the Green function $G(x)$ that connects two space-time points. In the EM case, we can proceed similarly and derive Maxwell equations from a propagation equation \cite{supp}, which is akin to Eq.~(\ref{jj}):
\begin{equation}\label{jj2}
  A^{\mu}(x)=(4\pi/c) \int G(x-x^{\prime})j_{(s)}^{\mu}(x^{\prime})d^{4}x^{\prime}.
\end{equation}
$A^{\mu}$ is in the Lorentz gauge ($\partial_{\mu}A^{\mu}=0$), as a consequence of charge conservation ($\partial_{\mu} j_{(s)}^{\mu}=0$) \cite{supp}. In the EM case, one assumes that the source current $j_{(s)}^{\mu}(x^{\prime})$ acts on a distant current $j^{\mu}(x)$ via the ``mediator'' $A^{\mu}(x)$, which couples to $j^{\mu}(x)$ through the term $A_{\mu}(x)j^{\mu}(x)$ in the corresponding Lagrangian that describes the dynamics of $j^{\mu}(x)$. We notice that, in contrast to Maxwell equations for the EM field, in Eq.~(\ref{max}) the current-density $\pi^{\mu}$ enters both sides of the equation. In the EM case though, something similar occurs when dealing with the differential equation that follows from Eq.~(\ref{wolf}) and the differential equation that
$G(\boldsymbol{\rho},z;\omega)$ satisfies.

We note in passing that Eq.~(\ref{kg2}) also holds for each component of $K^{\alpha \beta}$. Indeed, from Eq. (\ref{max2}) we get
    $\partial_{\alpha}\partial^{\alpha}K^{\beta\gamma}+\partial^{\beta}\left(-\partial_{\alpha}K^{\alpha \gamma}\right)+\partial^{\gamma}\left(\partial_{\alpha}K^{\alpha\beta}\right)=0$.
On view of Eqs.~(\ref{tensork}) and (\ref{max}),
\begin{equation}\label{kgmax}
  \square K^{\beta\gamma}=\kappa \, K^{\beta\gamma}.
\end{equation}
One can then show \cite{supp} that
\begin{equation}\label{k2}
  \kappa=
-\frac{1}{2}\frac{\int_{V}K_{\sigma\mu}K^{\sigma\mu}dV}{\int_{V}\pi_{\mu}\pi^{\mu}dV}.
\end{equation}
The quantity $K_{\sigma\mu}K^{\sigma\mu}$ is formally identical to the EM expression $F_{\sigma\mu}F^{\sigma\mu}$, which reads $\mathbf{E}^2-\mathbf{B}^2$ when written in terms of the electric and magnetic field vectors. Analogous vectors can be introduced, associated to $K_{\mu\nu}$. This suggests classifying the velocity fields in purely ``electric'' and purely ``magnetic''. They should have distinctive physical properties, according to $\kappa \gtrless 0$. 

The above results establish a parallelism with EM phenomena, so that diffraction, interference, etc., may take place also with respect to $K^{\alpha \beta}$.
Let us focus on interference patterns produced with massive particles. As shown in \cite{philippidis}, BT are obtained from a solution of the Helmholtz equation (\ref{helmholtz}) rather than from the Schr\"{o}dinger equation itself. Eq.~(\ref{helmholtz}) follows also from the wave equation. Hence, both BT and interference patterns can be explained in terms of Helmholtz's equation. Eq.~(\ref{kg6}) also leads to BT, by proceeding
as in the short-wave limit of optics. Let $u(x)$ stand for any of the $\pi^{\mu}$ in Eq.~(\ref{kg6}), and set
\begin{equation}\label{u}
u(x)=\rho^{1/2}(x) \exp(i \tilde{S}(x)/\lambda),
\end{equation}
where $\tilde{S}$ has the dimension of a length. On setting $u(x)$ in Eq.~(\ref{kg6}) and splitting real and imaginary parts, we get
\begin{eqnarray}
 \eta^{\mu\nu}\partial _{\mu}\tilde{S} \, \partial _{\nu}\tilde{S}-1-\lambda^2 \frac{\square \rho^{1/2}}{\rho^{1/2}} &=& 0, \label{shortwaver}\\
  2\eta^{\mu\nu}\partial _{\mu}\tilde{S} \, \partial _{\nu}\rho^{1/2}+\rho^{1/2}\square \tilde{S} &=& 0. \label{shortwavei}
\end{eqnarray}
The equations at zeroth- and first-order in $\lambda$ are
\begin{equation}\label{hh}
\eta^{\mu\nu}\partial _{\mu}\tilde{S}(x) \, \partial _{\nu}\tilde{S}(x)=1, \quad \partial_{\nu}\left(\rho(x) \, \partial^{\nu}\tilde{S}(x)\right)=0.
\end{equation}
These are, respectively, the (normalized) Hamilton-Jacobi equation
and the continuity equation for the probability current $\pi^{\nu}=\rho \, \partial^{\nu}\tilde{S}$.
The length $\lambda$ sets the scale for particle features to appear.

\subsection*{Bohmian trajectories}
For the Young setup, we can make a paraxial approximation and use two Gaussian beams propagating along the $z$ direction of the $XZ$-plane that contains the BT. We therefore set $u(t,x,z)=v(x,z)\exp[i(kz-\omega t)]$, with
\begin{equation}\label{v}
            \begin{aligned}
    v(x,z)= \frac{W_0}{W(z)}\left\{\exp\left[-\frac{(x-a)^2}{W^2(z)}\right]\exp\left[ik\frac{(x-a)^2}{2R(z)}-i\zeta(z)\right] \right.  \\
     \left. + \exp\left[-\frac{(x+a)^2}{W^2(z)}\right]\exp\left[ik\frac{(x+a)^2}{2R(z)}-i\zeta(z)\right]\right\}.
    \end{aligned}
    \end{equation}
Here, $W_0=(2z_0/k)^{1/2}$ is the waist radius and $z_0$ the Rayleigh range, while $W(z)=W_0(1+z^2/z_0^2)^{1/2}$, $R(z)=z[1+z^2/z_0^2]$ and $\zeta(z)=\tan^{-1}(z/z_0)$. The slits separation is $2a$. By writing $u(t,x,z)=v(x,z)\exp[i(kz-\omega t)]$ in polar form, see Eq.~(\ref{u}), we get $\rho(x,z)$ and $\tilde{S}(x,z)$. BT are integral curves of $v_i=\partial_i\tilde{S}$, $i=x,z$.

\begin{figure}[h!]
\centering
\fbox{\includegraphics[width=\linewidth]{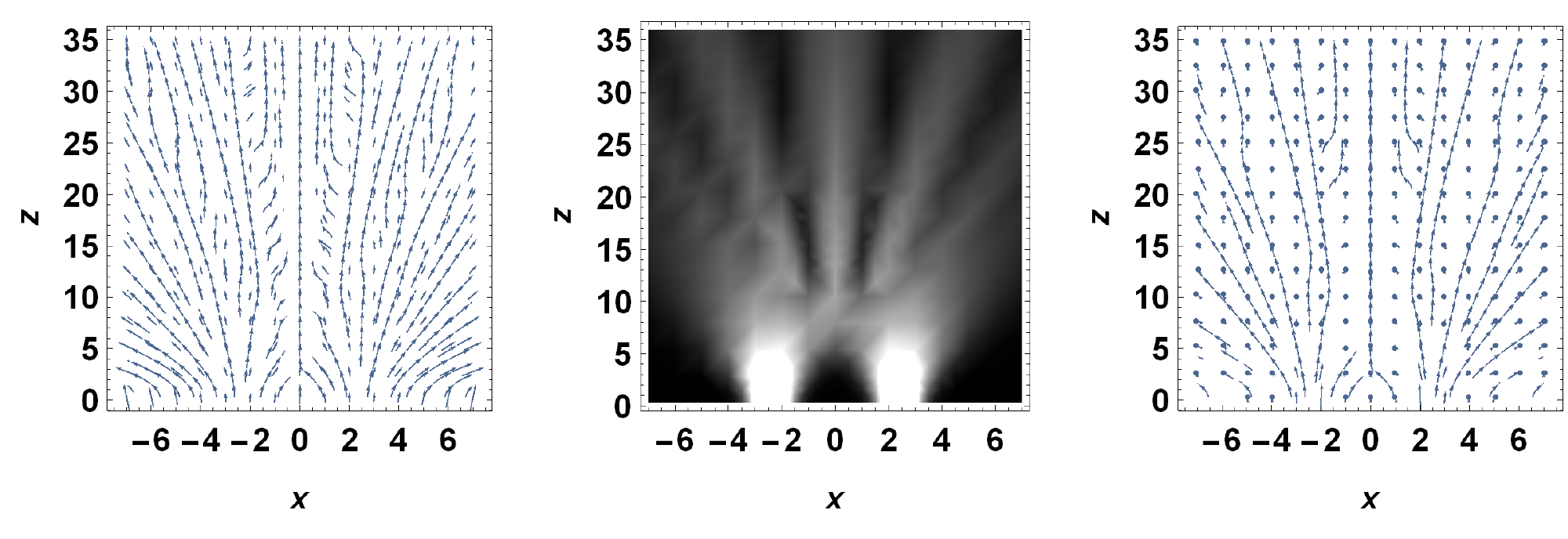}}
\caption{Bohmian trajectories in a Young setup with two Gaussian slits. Middle panel shows the probability density $\rho^{1/2}(x,z)$. Right panel contains the same trajectories as in the left panel, but weighted with $\rho^{1/2}$.}
\label{figp}
\end{figure}
The middle panel of Fig.~(\ref{figp}) shows $\rho^{1/2}$. The left panel shows the field $(v_x,v_z)$ and the right panel shows the weighted field $(\rho^{1/2}v_x,\rho^{1/2}v_z)$. This illustrates the effect of weighting with $\rho$ the integral curves of the field, thereby selecting from the infinity many ones (symbolically, those in the left panel) the cases actually realized in any experiment. We can only probabilistically assess which ones are these curves. They could be observed only as averaged trajectories, as it occurs when using weak-value measurements.
The parameters used in Fig.~(\ref{figp}) were chosen for illustrative purposes. Similar images have been obtained by Bliokh \textit{et al.} \cite{bliokh} with parameters taken from the experiments of Kocsis \textit{et al.} \cite{kocsis2011}. Another option would be to proceed as Philippidis \textit{et al.} did \cite{philippidis}, using Feynman's path integral method to get $\tilde{S}$, from which one can obtain BT.

\section*{Closing remarks}
Finally, let us say that while this work leaves many questions still open, it has reached its goal of showing that WPD and BT can fit within a fully classical framework.
The dynamical role assigned to EM vacuum should be accepted no more reluctantly than the analogous role of ``space-time'' in gravitation theory. If one is ready to accept that ``space-time curvature'' causes free particles to follow curvilinear trajectories, then one should also be ready to assign a similar role to EM vacuum. The latter has even more physical attributes, viz., $\epsilon_0$ and $\mu_0$, than the purely abstract concept that we call ``space-time''. The second model we have presented is such, that any objection one could raise against it, would most likely also apply to classical electrodynamics. On the other hand, we stress that our approach, while being physically motivated, should be given a sound mathematical basis. The so-called ``calculus of variations in the large'' could be an appropriate tool. In any case, we can envision a wide, uncharted territory in classical physics, which remains open to be explored.




\newpage

\onecolumngrid
\appendix

\begin{center}

\section*{Supplemental Information}

\end{center}

\subsection*{Carath\'{e}odory's formulation}

The usual approach in physics is to focus on a single
curve $\mathcal{C}$ that renders the action $I$ extremal: $\delta I \equiv \delta \int L(x,\dot{x})ds=0$. However, as the calculus of variations shows \cite{bliss,Caratheodory,Rund}, $\mathcal{C}$ exists only if
it can be embedded in a whole field of extremals, also known as a Mayer field. The existence of such a field implies some integrability conditions. Carath\'{e}odory's formulation \cite{Caratheodory}
makes clear how these integrability conditions relate to the Euler-Lagrange equations.

Let us summarize Carath\'{e}odory's approach. We adopt a relativistic formulation just for the sake of generality. A non-relativistic formulation could be established along similar lines. Extremals satisfying $\delta \int L(x,\dot{x})ds=0$ are the same as those
satisfying the so-called ``equivalent variational
problem'' \cite{Caratheodory} $\delta \int (L(x,\dot{x})-\dot{x}^{\mu}\partial _{\mu }S(x))ds=0$. Here, $S(x)$ is an auxiliary function. With its help, instead of seeking for a curve $\mathcal{C}$ that renders $I$ extremal, we seek for local extremal values. To this end, the Lagrangian is considered to be a function of $x$ and the velocity-field $v(x)$, i.e., $L=L(x,v(x))$. The extremal curve $\mathcal{C}$ is an integral curve of $v(x)$, i.e., it can be obtained by solving the first-order differential equations
\begin{equation}\label{aic}
  \frac{dx^{\mu }(s)}{ds}=v^{\mu}(x(s)).
\end{equation}
We thus seek for a velocity-field $v(x)$ whose integral curves render $\int Lds$ extremal. To this end, we impose the following condition:
\begin{equation}\label{a1}
 L(x,v)-v^{\mu }(x)\partial_{\mu }S(x)=0,
\end{equation}
and require that the expression on the left-hand side has zero as a stationary value with respect to variations of $v$. In the case of a maximum, for example, $L(x,w)-w^{\mu }\partial_{\mu }S<0$ for any field $w \neq v$. This guarantees that $\delta \int (L-\dot{x}^{\mu }\partial _{\mu }S)ds=0$, and so also $\delta \int L(x,\dot{x})ds=0$, because of the aforementioned equivalence of the two variational problems. Stationarity of the left-hand side of Eq.~(\ref{a1}) with respect to $v$ leads to
\begin{equation}\label{a2}
  \frac{\partial L(x,v)}{\partial v^{\mu}}=\frac{\partial S}{\partial x^{\mu}}.
\end{equation}
Eqs.~(\ref{a1}) and (\ref{a2}) are known as the fundamental equations of Carath\'{e}odory's approach.
On account of the integrability conditions $\partial ^{2}S/\partial x^{\mu }\partial x^{\nu }=\partial
^{2}S/\partial x^{\nu }\partial x^{\mu }$, it follows from Eq.~(\ref{a2}) that
\begin{equation}
\frac{\partial }{\partial x^{\mu }}\left( \frac{\partial L(x,v(x))}{\partial v^{\nu }}%
\right) -\frac{\partial }{\partial x^{\nu }}\left( \frac{\partial L(x,v(x))}{\partial
v^{\mu }}\right) =0.  \label{aintegrability}
\end{equation}%
From Eq.~(\ref{aintegrability}), we can get the equations of motion.
Indeed, from Eq.~(\ref{a1}) we obtain, by deriving with respect
to $x^{\mu}$,
\begin{equation}
\frac{\partial L}{\partial x^{\mu}}+\frac{\partial L}{\partial v^{\sigma}}%
\frac{\partial v^{\sigma}}{\partial x^{\mu}}=\frac{\partial v^{\sigma}}{%
\partial x^{\mu}}\frac{\partial S}{\partial x^{\sigma}}+v^{\sigma}\frac{%
\partial^{2}S}{\partial x^{\mu}\partial x^{\sigma}}.  \label{der}
\end{equation}
On using Eq.~(\ref{a2}), Eq.~(\ref{der}) reduces to
\begin{equation}
\frac{\partial L}{\partial x^{\mu}}=v^{\sigma}\frac{\partial^{2}S}{\partial
x^{\mu}\partial x^{\sigma}}. \label{dou}
\end{equation}
Using  $\partial ^{2}S/\partial x^{\mu }\partial x^{\nu }=\partial
^{2}S/\partial x^{\nu }\partial x^{\mu }$ first and then Eq.~(\ref{a2}), we get
\begin{equation}
\frac{\partial^{2}S}{\partial x^{\mu}\partial x^{\sigma}}=\frac{\partial^{2}S%
}{\partial x^{\sigma}\partial x^{\mu}}=\frac{\partial^{2}L}{\partial
x^{\sigma}\partial v^{\mu}}+\frac{\partial^{2}L}{\partial v^{\tau}\partial
v^{\mu}}\frac{\partial v^{\tau}}{\partial x^{\sigma}},  \label{double}
\end{equation}
so that Eq.~(\ref{dou}) reads
\begin{equation}
\frac{\partial L}{\partial x^{\mu}}=v^{\sigma}\frac{\partial^{2}L}{\partial
x^{\sigma}\partial v^{\mu}}+\frac{\partial^{2}L}{\partial v^{\tau}\partial
v^{\mu}}\frac{\partial v^{\tau}}{\partial x^{\sigma}}v^{\sigma}.
\label{devl}
\end{equation}
If we now evaluate this last relation along a single extremal, $%
dx^{\mu}/ds=v^{\mu}(x(s))$, we obtain, after recognizing the right
hand side of Eq.~(\ref{devl}) as $d(\partial L/\partial v^{\mu})/ds$, the
Euler-Lagrange equation:
\begin{equation}
\frac{d}{ds}\left( \frac{\partial L}{%
\partial\dot{x}^{\mu}}\right)-\frac{\partial L}{\partial x^{\mu}}=0 .  \label{e-l-r}
\end{equation}
Eq.~(\ref{devl}) is therefore more general than Eq.~(\ref{e-l-r}), which
follows from Eq.~(\ref{devl}), but not the other way around.

\subsection*{Hamilton-Jacobi equation}

The relativistic, free-particle Lagrangian reads
\begin{equation}
L=m c\left(\eta_{\mu\nu}v^{\mu}v^{\nu}\right)^{1/2}\equiv m c \, \phi.
\label{lagrangian2}
\end{equation}
From $\partial S/\partial x^{\mu}=\partial L/\partial v^{\mu}$ and Eq.~(\ref{lagrangian2}) we get
\begin{equation}\label{lc}
    v_{\mu}=\frac{\phi}{mc} \frac{\partial S}{\partial x^{\mu}}.
\end{equation}
On replacing Eq.~(\ref{lc}) in $\eta^{\mu\nu}v_{%
\mu }v_{\nu}=\phi^{2}$, we get the Hamilton-Jacobi equation
\begin{equation}
\eta^{\mu\nu}\left(\frac{\partial S}{\partial x^{\mu}}\right)\left(\frac{%
\partial S}{\partial x^{\nu}}\right)=m^{2}c^{2},  \label{hj3}
\end{equation}
without having introduced a Hamiltonian, which in the homogeneous case is a rather laborious task \cite{Caratheodory,Rund}.

\subsection*{Invariance of Carath\'{e}odory's fundamental equations under changes of the velocity field}

As we said before, Carath\'{e}odory's approach is based on two
fundamental equations: (\ref{a1}) and (\ref{a2}).
We deal with a relativistic, Lorentz invariant Lagrangian. Relativistic Lagrangians are homogeneous of the first degree in the velocities: $L(x,\lambda v)=\lambda L(x,v)$ (for $\lambda >0$). This property makes $\int L(x,\dot{x})ds$ invariant under parameter changes $s\rightarrow s^{\prime}$, a condition that must be met because $s$ has no physical meaning and can be arbitrarily chosen. We can then choose $s$ so that, say, $(\eta_{\mu\nu}\dot{x}^{\mu}\dot{x}^{\nu})^{1/2}=1$ along the sought-after extremal curve. We have a corresponding invariance when dealing with the velocity field $v^{\mu}(x)$. This time it is an invariance of the fundamental equations (\ref{a1}) and (\ref{a2}). Indeed,
multiplication of Eq.~(\ref{a1}) by a scalar function $\tilde{\phi}(x)>0$ leads to
\begin{equation}
\tilde{\phi}(x)\left( L(x,v)-v^{\mu }\partial _{\mu }S \right) =L(x,\tilde{\phi} v)-\left(
\tilde{\phi}\, v^{\mu }\right) \partial _{\mu }S=L(x,\tilde{v})-\tilde{v}^{\mu }\partial _{\mu }S,
\label{s2b}
\end{equation}%
with $\tilde{v}:=\tilde{\phi}\, v$.
From Eqs.~(\ref{a1}) and
(\ref{s2b}), it follows that
\[L(x,\tilde{v})-\tilde{v}^{\mu }\partial _{\mu }S(x) =0,\]which is Carath\'{e}odory's first fundamental equation for $\tilde{v}$.
Eq.~(\ref{a1}) defines the Lagrangian of the
``equivalent variational problem'': \[L^{\ast }(x,v):=L(x,v)-v^{\mu }\partial _{\mu }S.\]  We see that
\begin{equation}
\frac{\partial L^{\ast }(x,\tilde{v})}{\partial v^{\mu }}=\frac{\partial L^{\ast
}(x,\tilde{v})}{\partial \tilde{v}^{\nu }}\frac{\partial \tilde{v}^{\nu }}{\partial v^{\mu }}=\frac{%
\partial L^{\ast }(x,\tilde{v})}{\partial \tilde{v}^{\nu }}\left(\tilde{\phi}\,\delta^{\nu}_{\mu}\right)=\frac{%
\partial L^{\ast }(x,\tilde{v})}{\partial \tilde{v}^{\mu }} \, \tilde{\phi}.  \label{3}
\end{equation}%
On the other hand,
\begin{equation}
\frac{\partial L^{\ast }(x,\tilde{v})}{\partial v^{\mu }}=\frac{\partial }{\partial
v^{\mu }}\left( \tilde{\phi} L^{\ast }(x,v)\right) =\tilde{\phi}\, \frac{\partial L^{\ast
}(x,v)}{\partial v^{\mu }}=\tilde{\phi}\, \left( \frac{\partial L}{\partial v^{\mu }}-%
\frac{\partial S}{\partial x^{\mu }}\right) =0,  \label{4}
\end{equation}%
on account of Eq.~(\ref{a2}). Equations (\ref{3}) and (\ref{4}) imply that
$\partial L^{\ast }(x,\tilde{v})/\partial \tilde{v}^{\mu }=0$, i.e.,
\begin{equation}\label{few2}
    \frac{\partial L(x,\tilde{v})}{\partial \tilde{v}^{\mu }}=\frac{\partial S(x)}{\partial x^{\mu }},
\end{equation}
which is Carath\'{e}odory's second fundamental equation for $\tilde{v}$.
In summary, Eqs.~(\ref{a1}) and (\ref{a2}) hold if we replace $v$ by $\tilde{v}$. In other words, both velocity fields $v$ and $\tilde{\phi} \, v$ solve our variational problem for the same $S(x)$. This allows us to choose $\tilde{\phi}$ conveniently, e.g., such that $(\tilde{v}_{\nu}\tilde{v}^{\nu})^{1/2}=1$.

\subsection*{Velocity field $v^{\nu}$ for a given $\pi^{\nu}$}

As explained in the main text, given $\pi^{\nu}$, we can obtain $v^{\nu}$ from $n_0c\,v^{\nu}=(v_{\sigma}v^{\sigma})^{1/2}\pi^{\nu}$, with $\nu=0,\dots,3$. By squaring each of these four equations, we get
\begin{equation}
  n_0^2 c^2(v^{\nu})^2 = \left[(v^{0})^2-\mathbf{v}^2\right](\pi^{\nu})^2, \quad \nu=0,\dots,3.
\end{equation}
Written in matrix form, this system of equations for the $(v^{\nu})^2$ reads
\begin{equation}\label{vf1}M\left((v^{0})^2,(v^{1})^2,(v^{2})^2,(v^{3})^2\right)^T=0,
\end{equation}
where $T$ stands for transpose and
\begin{equation}\label{vf2}
 M= \left(
\begin{array}{cccc}
 n_0^2 c^2-\pi_0^2 & \pi_0^2 & \pi_0^2 & \pi_0^2 \\
 -\pi_1^2 & n_0^2 c^2+\pi_1^2 & \pi_1^2 & \pi_1^2 \\
 -\pi_2^2 & \pi_2^2 & n_0^2 c^2+\pi_2^2 & \pi_2^2 \\
 -\pi_3^2 & \pi_3^2 & \pi_3^2 & n_0^2 c^2+\pi_3^2 \\
\end{array}
\right).
\end{equation}
For the system (\ref{vf1}) to have a non-trivial solution, $\det M =0$. In our case,
\begin{equation}\label{vf3}
  \det M = (n_0 c)^6[(n_0 c)^2-\pi_{\nu}\pi^{\nu}].
\end{equation}
Hence, with $\pi_{\nu}\pi^{\nu}=n_0^2 c^2$, there are infinitely many solutions $(v^{\nu})^2$, in which one of the $v^{\nu}$ is a free parameter.

\subsection*{Two expressions for $\kappa$}

In the main text, we derived the equation
\begin{equation}\label{akg2}
  \square \pi^{\mu}=\kappa \pi^{\mu},
\end{equation}
from which it follows that $\pi_{\mu}\square \pi^{\mu}=\kappa \,\pi_{\mu} \pi^{\mu}$. By adding and subtracting $(\partial_{\sigma}\pi_{\mu})(\partial^{\sigma}\pi^{\mu})$ to $\pi_{\mu}\square \pi^{\mu}$, we obtain
$$(\partial_{\sigma}\pi_{\mu})(\partial^{\sigma}\pi^{\mu})+\pi_{\mu}\square \pi^{\mu}-(\partial_{\sigma}\pi_{\mu})(\partial^{\sigma}\pi^{\mu})=\partial_{\sigma}(\pi_{\mu}\partial^{\sigma}\pi{\mu})-
(\partial_{\sigma}\pi_{\mu})(\partial^{\sigma}\pi^{\mu}).$$
Hence,
\begin{equation}\label{ff1}
  \partial_{\sigma}(\pi_{\mu}\partial^{\sigma}\pi^{\mu})-
(\partial_{\sigma}\pi_{\mu})(\partial^{\sigma}\pi^{\mu})=\kappa \, \pi_{\mu}\pi^{\mu}.
\end{equation}
Integrating over a volume $V$ with boundary $S$, using the divergence theorem and assuming von Neumann or Dirichlet boundary conditions, we obtain
\begin{equation}\label{ff2}
  \int_{V}\partial_{\sigma}(\pi_{\mu}\partial^{\sigma}\pi^{\mu})\,dV=\int_{S}(\pi_{\mu}\partial^{\sigma}\pi^{\mu})dS_{\sigma}=0.
\end{equation}
Thus, integrating over $V$ both sides of Eq.~(\ref{ff1}), we get
\begin{equation}\label{ak1}
  \kappa=-\frac{\int_{V}(\partial_{\sigma}\pi_{\mu})(\partial^{\sigma}\pi^{\mu})dV}{\int_{V}\pi_{\mu}\pi^{\mu}dV}.
\end{equation}

We now consider the tensor
\begin{equation}\label{atensork}
K^{\alpha \beta}=\partial^{\alpha}\pi^{\beta}-\partial^{\beta}\pi^{\alpha}.
\end{equation}
We see that
$$K_{\sigma\mu}K^{\sigma\mu}=(\partial_{\sigma}\pi_{\mu}-
\partial_{\mu}\pi_{\sigma})(\partial^{\sigma}\pi^{\mu}-\partial^{\mu}\pi^{\sigma})=
2\left[(\partial_{\sigma}\pi_{\mu})(\partial^{\sigma}\pi^{\mu})-(\partial_{\sigma}\pi_{\mu})(\partial^{\mu}\pi^{\sigma})\right].$$
Hence,
\begin{equation}\label{ff3}
(\partial_{\sigma}\pi_{\mu})(\partial^{\sigma}\pi^{\mu})=\frac{1}{2}K_{\sigma\mu}K^{\sigma\mu}+(\partial_{\sigma}\pi_{\mu})(\partial^{\mu}\pi^{\sigma}).
\end{equation}
On the other hand, using $\partial_{\sigma}\pi^{\sigma}=0$, we obtain
\begin{equation}\label{ff4}
\partial_{\sigma}\left[\pi_{\mu}\partial^{\mu}\pi^{\sigma}\right]=(\partial_{\sigma}\pi_{\mu})(\partial^{\mu}\pi^{\sigma})
+\pi_{\mu}\partial^{\mu}(\partial_{\sigma}\pi^{\sigma})=(\partial_{\sigma}\pi_{\mu})(\partial^{\mu}\pi^{\sigma}).
\end{equation}
From Eqs.~(\ref{ff3}) and (\ref{ff4}), we derive
\begin{equation}\label{ff5}
\int_V (\partial_{\sigma}\pi_{\mu})(\partial^{\sigma}\pi^{\mu})dV=\frac{1}{2}\int_{V} K_{\sigma\mu}K^{\sigma\mu}dV+\int_V \partial_{\sigma}\left[\pi_{\mu}\partial^{\mu}\pi^{\sigma}\right]dV=\frac{1}{2}\int_{V} K_{\sigma\mu}K^{\sigma\mu}dV,
\end{equation}
where we have used that $\int_V \partial_{\sigma}\left[\pi_{\mu}\partial^{\mu}\pi^{\sigma}\right]dV=\int_S \pi_{\mu}\partial^{\mu}\pi^{\sigma}dS_{\sigma}=0$. Eqs.~(\ref{ak1}) and (\ref{ff5}) then lead to
\begin{equation}\label{k2}
  \kappa=
-\frac{1}{2}\frac{\int_{V}K_{\sigma\mu}K^{\sigma\mu}dV}{\int_{V}\pi_{\mu}\pi^{\mu}dV}.
\end{equation}

\subsection*{Maxwell equations}

In the main text, we considered the propagation equation
\begin{equation}\label{ajj2}
  A^{\mu}(x)=\frac{4\pi}{c} \int G(x-x^{\prime})j_{(s)}^{\mu}(x^{\prime})d^{4}x^{\prime}.
\end{equation}
From this equation, it follows that
\begin{eqnarray}\label{ajj3}
  \partial_{\mu} A^{\mu}(x)&=&\frac{4\pi}{c} \int \left(\partial_{\mu}G(x-x^{\prime})\right)j_{(s)}^{\mu}(x^{\prime})d^{4}x^{\prime} \nonumber \\
  &=&\frac{4\pi}{c} \int  \left(-\partial_{\mu}^{\prime}G(x-x^{\prime})\right)j_{(s)}^{\mu}(x^{\prime})d^{4}x^{\prime}
   \nonumber \\
  &=&-\frac{4\pi}{c} \int \partial_{\mu}^{\prime}\left(G(x-x^{\prime})j_{(s)}^{\mu}(x^{\prime})\right)d^{4}x^{\prime}+
  \frac{4\pi}{c} \int G(x-x^{\prime})\left(\partial_{\mu}^{\prime}j_{(s)}^{\mu}(x^{\prime})\right)d^{4}x^{\prime}=0,
\end{eqnarray}
where in the last equality the first integral is shown to vanish by applying Gauss' theorem and the boundary condition $j_{\mu} \rightarrow 0$ at infinity, and the second integral vanishes on account of charge conservation:
$\partial_{\mu}j_{(s)}^{\mu}=0$.

We have then that, as a consequence of Eqs.~(\ref{ajj2}) and (\ref{ajj3}),
\begin{equation}\label{an3}
  \square A^{\nu}=\frac{4\pi}{c} j_{(s)}^{\nu}, \quad \text{and} \quad \partial_{\nu}A^{\nu}=0.
\end{equation}
From the above equations, it immediately follows that $F^{\mu \nu}=\partial^{\mu}A^{\nu}-\partial^{\nu}A^{\mu}$ satisfies the Maxwell equations
\begin{eqnarray}
    \partial _{\mu }F^{\mu \nu}&=&\frac{4\pi}{c} j_{(s)}^{\nu}, \label{an5a} \\
     \partial_{\alpha}F_{\beta\gamma}+\partial_{\beta}F_{\gamma \alpha}+\partial_{\gamma}F_{\alpha\beta}&=&0, \label{an5b}
\end{eqnarray}
where (\ref{an5b}) is an identity that follows from the definition of $F^{\mu \nu}$. Hence, the essential thrust of Maxwell equations lies on the propagation equation (\ref{ajj2}). Of course, (\ref{an5a}) and (\ref{an5b}) are gauge invariant whereas
(\ref{ajj2}) is not. Classically, gauge invariance is relevant only in relation to the coupling of $A^{\mu}$ to some charge through a term of the form $(e/c)A_{\mu}v^{\mu}$ in the Lagrangian. As the charge's equations of motion depend on $A^{\mu}$ only through $F^{\mu\nu}$, one may say that not $A^{\mu}$ itself but $F^{\mu\nu}$ is physically meaningful. It is easy to see that a gauge transformation $A^{\mu} \rightarrow \tilde{A}^{\mu}=A^{\mu}+\partial^{\mu} W$ amounts to a change $S \rightarrow \tilde{S}=S+(e/c)W$ in the auxiliary function used in Carath\'{e}odory's fundamental equations (\ref{a1}) and (\ref{a2}). Thus, it does not matter that $A^{\mu}$ is in the Lorentz gauge.
The only way to observe $A^{\mu}$ is by coupling it to a (test) charge, in order to register the latter's response to it. This response is gauge invariant, according to (\ref{a1}) and (\ref{a2}), which imply the Euler-Lagrange equations.

\end{document}